\documentclass[aps,prx,superscriptaddress,amssymb,showpacs,onecolumn,10pt]{revtex4-1}
\usepackage{amsfonts} 
\usepackage{amsmath}
\usepackage{amssymb}
\usepackage{graphicx}
\usepackage{hyperref}
\usepackage{color}
\usepackage{natbib}  
\usepackage{bm} 
\usepackage{isomath} 
\usepackage{braket}
\usepackage{subfigure}

\input{epsf}
\newcommand{\ignore}[1]{}
\newcommand{\nobibentry}[1]{{\let\nocite\ignore\bibentry{#1}}}

\begin{document}

\title{Performance of continuous quantum thermal devices indirectly connected to environments}

\author{J. Onam Gonz{\'a}lez}
%\affiliation{IUdEA Instituto Universitario de Estudios Avanzados, Universidad de La Laguna, 38203 Spain}
%\affiliation{Dpto. F\'{\i}sica Universidad de La Laguna, 38203 Spain}

\author{Daniel Alonso}
%\affiliation{IUdEA Instituto Universitario de Estudios Avanzados, Universidad de La Laguna, 38203 Spain}
%\affiliation{Dpto. F\'{\i}sica Universidad de La Laguna, 38203 Spain}

\author{Jos\'{e} P. Palao}
\affiliation{IUdEA Instituto Universitario de Estudios Avanzados, Universidad de La Laguna, 38203 Spain}
\affiliation{Dpto. F\'{\i}sica Universidad de La Laguna, 38203 Spain}


\begin{abstract}

A general quantum thermodynamics network is composed of thermal devices connected to the environments through quantum wires. The coupling between the devices and the wires may introduce additional decay channels which modify the system performance with respect to the directly-coupled device. We analyze this effect in a quantum three-level device connected to a heat bath or to a work source through a two-level wire. The steady state heat currents are decomposed into the contributions of the set of simple circuits in the graph representing the master equation. Each circuit is associated with a mechanism in the device operation and the system performance can be described by a small number of circuit representatives of those mechanisms. Although in the limit of weak coupling between the device and the wire the new irreversible contributions can become small, they prevent the system from reaching the Carnot efficiency.
\end{abstract}

\maketitle
% Keywords
%\keyword{quantum thermodynamics; graph theory; thermodynamic performance}

%%%%%%%%%%%%%%%%%%%%%%%%%%%%%%%%%%%%%%%%%%
% Only for the journal Data:

%\dataset{DOI number or link to the deposited data set in cases where the data set is published or set to be published separately. If the data set is submitted and will be published as a supplement to this paper in the journal Data, this field will be filled by the editors of the journal. In this case, please make sure to submit the data set as a supplement when entering your manuscript into our manuscript editorial system.}

%\datasetlicense{license under which the data set is made available (CC0, CC-BY, CC-BY-SA, CC-BY-NC, etc.)}

%%%%%%%%%%%%%%%%%%%%%%%%%%%%%%%%%%%%%%%%%%
\section{Introduction}

Continuous quantum thermal devices are quantum systems connected to several baths at different temperatures and to work sources \cite{Kosloff2014}. Their operation is necessarily irreversible when the heat currents are non-negligible. One of the possible irreversible processes is the ubiquitous finite-rate heat transfer effect considered in endoreversible models. In these models the control parameters can be tunned to reach the reversible limit but at vanishing energy flows. Examples are the three-level and the two-qubit absorption refrigerators \cite{Palao2001,Linden2010,Levy2012,Correa2014c}. In other models, as the power-driven three-level maser \cite{Scovil1959,Kosloff2014} and the three-qubit absorption refrigerators \cite{Linden2010,Correa2013}, additional irreversible processes appear such as heat leaks and internal dissipation \cite{Kosloff2014,Correa2015}, which are detrimental to the device performance. {Several experimental realizations of these continuous quantum thermal devices have been proposed, for example nano-mechanical oscillators or atoms interacting with optical resonators} \cite{Mari2012,Mitchison2016}, {atoms interacting with nonequilibrium electromagnetic fields} \cite{Leggio2015}, {superconducting quantum interference devices} \cite{Chen2012}, {and quantum dots} \cite{Venturelli2013}. {Besides, the coupling between artificial atoms and harmonic oscillators is experimentally feasible nowadays} \cite{Peng2015}, {opening the perspective of connecting thermal devices to environments through quantum systems}. 

Although the most general design of a quantum thermal network is composed of thermal devices and wires \cite{Kosloff2013}, the device performance has been usually analyzed assuming a direct contact with the environments. The coupling between the device and a quantum probe has been suggested to characterize the device irreversible processes \cite{Correa2016}. In this paper we adopt a different perspective and study the additional irreversible processes induced by the coupling between the device and the wire. In particular we will analyze in detail the performance of a system composed of a three-level device and a two-level wire connected to a work bath or source. When the system is weakly coupled to the baths, its evolution is described by a quantum master equation in which the dynamics of the populations can be decoupled from the coherences choosing an appropriate basis \cite{Breuer2002}. This property implies the positivity of the entropy production along the system evolution \cite{Spohn1977,Alicki1979}, and is broken when some uncontrolled approximations are considered in the derivation of the quantum master equation \cite{Levy2014}. The Pauli master equation for the populations (in the following simply the master equation) is a particular example of the general master equations considered in stochastic thermodynamics for systems connected to multiple reservoirs  \cite{Broek2015}. For long enough time, the system reaches a non-equilibrium steady state where the heat currents $\dot{\mathcal{Q}}_\alpha$ describe the energy transfer between the system and the baths, and the power $\mathcal{P}$ characterizes the energy exchange with the work source \cite{Kosloff2013}. Although the heat currents can be obtained directly from the steady solution of the master equation, to identify the different irreversible processes contributing to them is in general a very complicated task.

An alternative approach to analyze non-equilibrium processes is the decomposition of the steady state fluxes and the entropy production in the contribution due to simple circuits \cite{Hill1966,Jiang2004,Kalpazidou2007}, fundamental circuits \cite{Schnakenberg1976,Andrieux2007,Polettini2015} or cycles \cite{Altaner2012} of the graph representation of the master equation. We will consider a decomposition in the full set of simple circuits that combined with the all minors matrix-tree theorem \cite{Moon1994} lead to very simple expressions for the steady state heat currents. More importantly, each circuit can be interpreted as a thermodynamically consistent unit and its contribution to the different irreversible processes can be easily identified \cite{Gonzalez2016}. Although the number of circuits may be very large, we will show that the system performance can be described by means of a reduced number of circuit representatives \cite{Knoch2015}.

The paper is organized as follows: Section \ref{sec:theory} presents a brief review of the derivation of the quantum master equation for a device coupled to a heat bath through a quantum wire. Next the graph representation of the master equation and the decomposition in simple circuits is discussed, with special emphasis on the characterization of the steady state heat currents. Some procedures to determine the set of simple circuits are described in Appendix A. The absorption refrigerator composed of a three-level device connected to a work bath through a two-level wire is studied in Section \ref{sec:bathsource} and some circuit representatives are suggested to describe the system performance. The same analysis is applied to a system driven by a periodic classical field in Section \ref{sec:worksource}, which includes a brief discussion of the derivation of the master equation for the time dependent Hamiltonian. In this case we study the performance operating as a refrigerator or as an engine. Finally, we draw our conclusions in Section \ref{sec:conclusions}.

%%%%%%%%%%%%%%%%%%%%%%%%%%%%%%%%%%%%%%%%%%
\section{Circuit decomposition of the steady state heat currents and entropy production}\label{sec:theory}

%%%%%%%
\subsection{The master equation}

We consider a system composed of a quantum thermal device with Hamiltonian $\hat{H}_{D}$  directly coupled with a cold bath and a hot bath, at temperatures $T_c$ and $T_h$, and a quantum wire with Hamiltonian $\hat{H}_{wire}$ which connects the device to an additional bath at temperature $T_w$ (work bath). The situation in which the system is driven instead by a work source will be discussed in section \ref{sec:worksource}. The total Hamiltonian reads

\begin{equation}
\hat{H}\,=\,\hat{H}_{D}\,+\,\hat{H}_{D,wire}\,+\,\hat{H}_{wire}
\,+\,\hat{H}_{wire,w}\,+\,\hat{H}_w\,+\,
\sum_{\alpha=c,h}\,\left(\hat{H}_{D,\alpha}\,+\,\hat{H}_\alpha\right)\,,
\end{equation}

where $\hat{H}_{D,wire}$ is the coupling between the device and the wire, $\hat{H}_{D,\alpha}$ and $\hat{H}_{wire,w}$ the coupling terms of the device and the wire with the baths, and $\hat{H}_\alpha$ the bath Hamiltonians. We assume that the coupling terms of the system with the baths are $\sqrt{\gamma_\alpha}\,\hbar\hat{S}^{\alpha}\otimes\hat{B}^\alpha$, where $\hat{S}^\alpha$ is a device or wire Hermitian operator, $\hat{B}^\alpha$ is a bath operator and $\gamma_\alpha$ characterizes the coupling strength.

If the system is weakly coupled with the baths and its relaxation time scale is slow compared with the correlation times of the baths, the system evolution can be described by a Markovian quantum master equation for its reduced density operator $\hat{\rho}$. The procedure to obtain this quantum master equation is described for example in \cite{Breuer2002}. Here we just comment on the final result. Let $\hat{U}_S(t)=\exp(-i\hat{H}_S t/\hbar)$ denotes the evolution operator corresponding to the system Hamiltonian $\hat{H}_S=\hat{H}_{D}\,+\,\hat{H}_{D,wire}\,+\,\hat{H}_{wire}$. The essential elements in the quantum master equation can be identified from the following decomposition of the operators $\hat{S}^\alpha$ in interaction picture

\begin{equation}
\hat{U}_S^\dagger(t)\,\hat{S}^{\alpha}\,\hat{U}_S(t)\,=\,\sum_{\omega>0}
\hat{S}_{\omega}^{\alpha}\exp(-i\omega t)\,+\,\hat{S}_{\omega}^{\alpha\dagger}\exp(i\omega t)\,,
\end{equation}

where $\sum_{\omega>0}$ denotes the summation over the positive transition frequencies {$\omega_{ij}=\omega_j-\omega_i$} between eigenstates of $\hat{H}_S$. The difference between the spectrum of $\hat{H}_S$ and $\hat{H}_{D}$ makes the frequencies and terms in the previous decomposition different from the one corresponding to the device directly coupled with the baths, and leads to new decay channels. This is the origin of the additional irreversible processes.

When system intrinsic dynamics is fast compared to the relaxation dynamics, the rotating wave approximation applies and the Lindbland-Gorini-Kossakovsky-Sudarshan (LGKS) generators of the irreversible dynamics associated with each bath can be written as 

\begin{equation}\label{eq:generators}
\mathcal{L}_\alpha[\hat{\rho}(t)] \,=\, \sum_{\omega>0}\;
\Gamma_{\omega}^\alpha\left(\hat{S}_{\omega}^{\alpha}\hat{\rho}\hat{S}_{\omega}^{\alpha\dagger}
-\frac{1}{2}\{\hat{S}_{\omega}^{\alpha\dagger}\hat{S}_{\omega}^{\alpha},\hat{\rho}\}\right)
\,+\,\Gamma_{-\omega}^\alpha\left(\hat{S}_{\omega}^{\alpha\dagger}\hat{\rho}\hat{S}_{\omega}^{\alpha}
-\frac{1}{2}\{\hat{S}_{\omega}^{\alpha}\hat{S}_{\omega}^{\alpha\dagger},\hat{\rho}\}\right)\,,
\end{equation}

We have introduced the anticommutators $\{\hat{S}\hat{S}^\dagger,\hat{\rho}\}=\hat{S}\hat{S}^\dagger\hat{\rho}+\hat{\rho}\hat{S}\hat{S}^\dagger$. In the following we will consider bosonic baths of physical dimensions $d_\alpha$ and coupling operators $\hat{B}^\alpha\propto\sum_\mu\sqrt{\omega_\mu}(\hat{b}_\mu^\alpha+\hat{b}_\mu^{\alpha\dagger})$. The summation is over all the bath modes of frequencies $\omega_\mu$ and annhilation operators $\hat{b}_\mu$. With this choice the rates $\Gamma^\alpha_{\pm\omega}$ are \cite{Breuer2002}
\begin{eqnarray}\label{eq:rates}
\Gamma^\alpha_{\omega} &=& \gamma_\alpha\,(\omega/\omega_0)^{d_\alpha}[N^\alpha(\omega)+1] \,,
\nonumber \\
\Gamma^\alpha_{-\omega} &=& \Gamma^\alpha_{\omega}\exp(-\omega\hbar/k_BT_\alpha) \,,
\end{eqnarray}

with $N^\alpha(\omega)=[\exp(\omega\hbar/k_BT_\alpha)-1]^{-1}\,$, $k_B$ the Boltzmann constant, and the frequency $\omega_0$ depends on the physical realization of the coupling with the bath. Finally, assuming that the Lamb shift of the unperturbed energy levels is small enough to be neglected, the quantum master equation in the Shr\"odinger picture is given by 

\begin{equation}\label{eq:quantummaster}
\frac{d}{dt} \hat{\rho}(t)\,=\,-\frac{i}{\hbar}[\hat{H}_S,\hat{\rho}(t)]\,+\,\sum_{\alpha=c,w,h}\mathcal{L}_\alpha[\hat{\rho}(t)]\,.
\end{equation}

This quantum master equation is in the standard Lindblad form and defines a generator of a dynamical semigroup. If the spectrum of $\hat{H_S}$ is non-degenerated, equation (\ref{eq:quantummaster}) for the populations of the $N$ eigenstates $|i\rangle$ of $\hat{H}_S$, $p_i=\langle i|\hat{\rho}|i\rangle$, reduces to \cite{Breuer2002}

\begin{equation}\label{eq:master}
\frac{d}{dt} p_i(t)\,=\,\sum_{j=1}^{N} \; \sum_{\alpha=c,w,h}\, W_{ij}^\alpha\, p_j(t)\,=\,\sum_{j=1}^{N} \, W_{ij}\, p_j(t)\,,
\end{equation}

where $W_{ij}^\alpha$ is the transition rate from the state $j$ to the state $i$ due to the coupling with the bath $\alpha$. In the following $\mathbf{W}$ will denote the matrix with elements $W_{ij}=\sum_{\alpha=c,w,h}W_{ij}^\alpha$. The diagonal elements satisfy

\begin{equation}\label{eq:diag}
W^\alpha_{ii}\,=\,-\sum_{j\neq i}W^\alpha_{ji}\,,
\end{equation}

implying the conservation of the normalization. Besides, as a consequence of the Kubo-Martin-Schwinger condition in (\ref{eq:rates}), the forward and backward transition rates are related by

\begin{equation}\label{eq:nondiag}
\frac{W_{ji}^\alpha}{W_{ij}^\alpha}\,=\,\exp\left(-\frac{\omega_{ij}\hbar}{k_BT_\alpha}\right)\,.
\end{equation}

Equation (\ref{eq:master}) is the starting point of our analysis. When the system is driven by a work source we will arrive to an equation with a similar structure, and the results described below will also apply to that case.

%%%%%%%
\subsection{Circuit fluxes and affinities}

Here we describe how to determine the heat currents $\dot{\mathcal{Q}}_\alpha$ and the entropy production $\dot{\mathcal{S}}$ in the steady state. In the following we assume that the currents are positive when the energy flows towards the system. The method is based on the representation of the master equation (\ref{eq:master}) by a connected graph $\mathcal{G}(V,E)$, being $|V|=N$ the number of vertices, representing the system states, and $|E|$ the number of undirected edges, representing the transitions between different states. A simple circuit $\mathcal{C}_\nu$ of $\mathcal{G}$ is a closed path with no repetition of vertices or edges. Some procedures to determine the set of circuits in a graph are discussed in appendix A. Each one of the two possible different orientations of a simple circuit, denoted by $\vec{\mathcal{C}}_\nu$ and $-\vec{\mathcal{C}}_\nu$, is a cycle. A cycle then consists of a sequence of directed edges with transition rates $W_{ij}^\alpha$, and it has an associated algebraic value \cite{Schnakenberg1976} 

\begin{equation}
\mathcal{A}(\vec{\mathcal{C}}_\nu)\,=\,\prod_{\alpha=c,w,h}\mathcal{A}^\alpha(\vec{\mathcal{C}}_\nu)\,,
\end{equation}

with

\begin{equation}
\mathcal{A}^\alpha(\vec{\mathcal{C}}_\nu)\,=\,\prod_{ij \in \nu} W_{ij}^\alpha\,,
\end{equation}

where $\prod_{ij \in \nu}$ denotes the product of all the transition rates due to the bath $\alpha$ in the cycle $\vec{\mathcal{C}}_\nu$. If the cycle does not involve the bath $\alpha$, $\mathcal{A}^\alpha(\vec{\mathcal{C}}_\nu)=1$ for consistency.

The cycle affinity \cite{Schnakenberg1976} is defined by 

\begin{equation}
X(\vec{\mathcal{C}}_\nu)\,=\,\sum_{\alpha=c,w,h}\,X^\alpha(\vec{\mathcal{C}}_\nu)
\,=\,k_B\,\ln\left(\frac{\mathcal{A}(\vec{\mathcal{C}}_\nu)}
{\mathcal{A}(-\vec{\mathcal{C}}_\nu)}\right)\,,
\end{equation}

where the affinity associated with each bath is $X^\alpha(\vec{\mathcal{C}}_\nu)=k_B\ln[{\mathcal{A}^\alpha(\vec{\mathcal{C}}_\nu)}/{\mathcal{A}^\alpha(-\vec{\mathcal{C}}_\nu)}]$. When the system is only coupled with thermal baths, the same amount of energy is taken and transferred to them in a complete cycle, implying $\sum_\alpha T_\alpha X^\alpha(\vec{\mathcal{C}}_\nu)=0$. However, when the system is in addition coupled to a work source the summation may differ from zero, indicating the net exchange of energy between the work source and the baths. The cycle flux is defined by \cite{Gonzalez2016}

\begin{equation}\label{eq:cycleflux}
I(\vec{\mathcal{C}}_\nu)=D^{-1}\det(-\mathbf{W}|\mathcal{C}_\nu)
[\mathcal{A}(\vec{\mathcal{C}}_\nu)\,-\,\mathcal{A}(-\vec{\mathcal{C}}_\nu)]\,,
\end{equation}

where $D\,=\,|\det(\widetilde{\mathbf{W}})|$. The matrix $\widetilde{\mathbf{W}}$ is obtained from the rate matrix ${\mathbf{W}}$ replacing the elements of an arbitrary row by ones, and $(-\mathbf{W}|\mathcal{C}_\nu)$ denotes the matrix resulting from removing from $-\mathbf{W}$ all the rows and columns corresponding to the vertices of the circuit $C_\nu$. Considering the relation between the diagonal and non-diagonal elements of $\mathbf{W}$, the determinant of $(-\mathbf{W}|C_\nu)$ is always positive. The opposite cycle affinities and flux change according to $X^{(\alpha)}(-\vec{\mathcal{C}}_\nu)=-X^{(\alpha)}(\vec{\mathcal{C}}_\nu)$ and $I(-\vec{\mathcal{C}}_\nu)=-I(\vec{\mathcal{C}}_\nu)$.

Using these definitions, the steady state heat current between the system and a bath associated with a simple circuit is

\begin{equation}\label{eq:circuitcurrent}
\dot{\mathcal{Q}}_\alpha(\mathcal{C}_\nu)
\,=\,-\,T_\alpha\,I(\vec{\mathcal{C}}_\nu)\,X^{\alpha}(\vec{\mathcal{C}}_\nu)\,,
\end{equation}

and the steady state entropy production is given by $\dot{\mathcal{S}}({C}_\nu)\,=\, I(\vec{\mathcal{C}}_\nu)\,X(\vec{\mathcal{C}}_\nu)$. At this point it is important to notice that although the affinity and the flux are defined for each cycle, the steady state heat currents and entropy production are independent of the cycle orientation and can then be assigned to the circuit without any ambiguity. Besides each circuit is consistent with the first and second law of thermodynamics as
$\sum_{\alpha=c,w,h}\dot{\mathcal{Q}}_\alpha(\mathcal{C}_\nu)=0$  and $\dot{S}({C}_\nu)\ge 0$ \cite{Hill1966,Kalpazidou2007,Jiang2004,Gonzalez2016}. Finally the heat currents and the entropy production in the steady state can be obtained by adding the contribution of all the simple circuits of the graph, $\dot{\mathcal{Q}}_\alpha=\sum_\nu\dot{\mathcal{Q}}_\alpha(\mathcal{C}_\nu)$ and $\dot{\mathcal{S}}=\sum_\nu\dot{\mathcal{S}}({C}_\nu)$.

The relative importance of the contribution due to a simple circuit to the heat current (\ref{eq:circuitcurrent}) is determined by both its affinity {$X^\alpha(\vec{\mathcal{C}}_\nu)$} and flux {$I(\vec{\mathcal{C}}_\nu)$} (\ref{eq:cycleflux}). When the system is coupled with thermal baths, the circuits can be classified as trivial circuits (all the affinities $X^\alpha=0$), circuits associated with heat leaks (one of the affinities is zero) and tricycles (the three affinities are different from zero) \cite{Gonzalez2016}. Trivial circuits do not contribute to the steady state heat currents or entropy production. Circuits associated with heat leaks only connect two baths, and the heat always flows {from} the higher to the lower temperature bath. Tricycles \cite{Kosloff2013} are circuits connecting the three baths, independently of the number of edges involved. When the system is coupled instead with a work source, the circuits associated with heat leaks are identified from the condition $T_cX^c+T_hX^h=0$ (as there is not net energy exchange with the source), and  the tricycles from $T_cX^c+T_hX^h\neq 0$ \cite{Gonzalez2016}.

The analysis of the circuit flux is more complicated as the term $\mathcal{A}(\vec{\mathcal{C}}_\nu)\,-\,\mathcal{A}(-\vec{\mathcal{C}}_\nu)$ strongly depends on the system parameters. In any case, the number of terms in the determinant of the matrix $(-\mathbf{W}|\mathcal{C}_\nu)$ decreases when the number of edges in the circuit increases. Then the non-trivial circuits with a lower number of edges are the dominant contribution to the heat currents in a large range of the parameters, and the system operation as a thermal machine is mainly determined by tricycles with a low number of edges.

%%%%%%%%%%%%%%%%%%%%%%%%%%%%%%%%%%%%%%%%%%
\section{Quantum three-level device coupled through a two-level system to a work bath}\label{sec:bathsource}

\begin{figure}[h]
\includegraphics[width=0.9\linewidth]{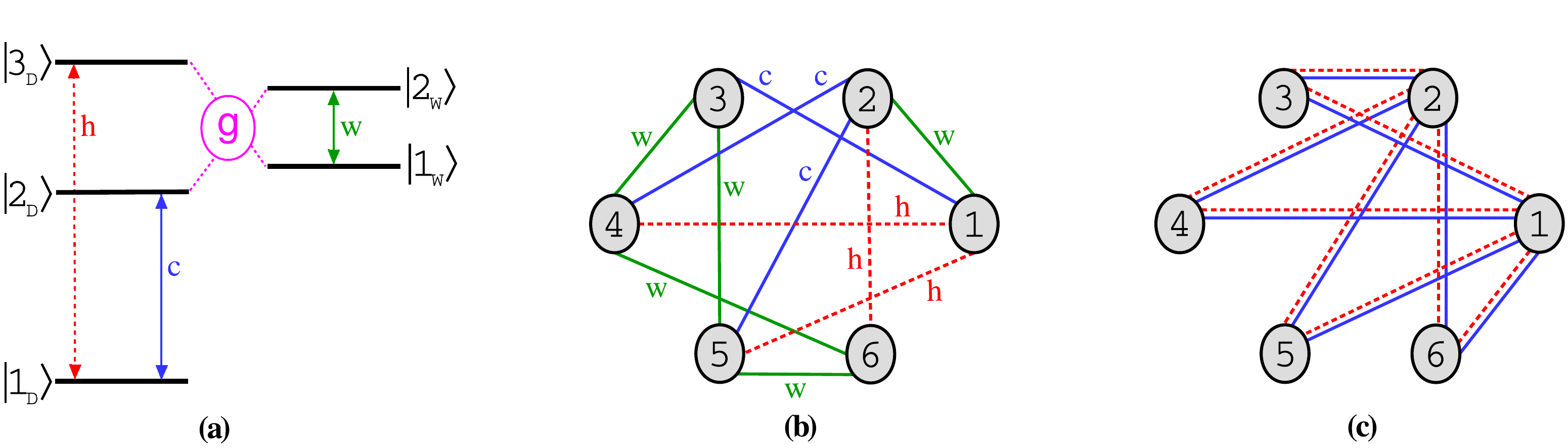}
\caption{(a) Schematic illustration of a three-level device coupled to a work bath or source through a two-level wire. (b) Graph representation of the master equation when the wire connects a work bath. The six vertices represent the eigenstates of $\hat{H}_S$ and the eleven edges the transitions assisted by the cold (blue lines), work (green lines) and hot (dashed red lines) baths, labeled by $c$, $w$ and $h$ respectively. (c) Graph representation of the master equation when the system is driven by a periodic classical field. Now the vertices correspond to the eigenstates of $\hat{H}_2$, and there are eight pairs of parallel edges associated with the cold (solid lines) and hot (dashed lines) baths.}\label{fig:figure1}
\end{figure} 

An absorption refrigerator is a thermal device extracting heat from a cold bath and rejecting it to a hot bath at rates $\dot{\mathcal{Q}}_c$ and $\dot{\mathcal{Q}}_h$ respectively. This process is assisted by the heat $\dot{\mathcal{Q}}_w$ extracted from a work bath at higher temperature. Its coefficient of performance (COP) is given by $\varepsilon=\dot{\mathcal{Q}}_c/\dot{\mathcal{Q}}_w$. The simplest model of quantum absorption refrigerator is a three-level system directly coupled with the heat baths \cite{Palao2001,Linden2010}. When $T_c<T_h<T_w$ the refrigerator operates in the cooling window $\omega_c<\omega_{c,rev}=\omega_hT_c(T_w-T_h)/[T_h(T_w-T_c)]$, with $\omega_c$, $\omega_h$ and $\omega_w=\omega_h-\omega_c$ the frequencies of the transitions coupled with the cold, hot and work baths respectively.  In the limit of $\omega_c$ approaching from below to $\omega_{c,rev}$ the COP reaches the Carnot limit $\varepsilon_C=T_c(T_w-T_h)/[T_w(T_h-T_c)]$, as the only source of irreversibility is the finite heat transfer rate through the thermal contacts. To analyze the effect of the indirect coupling we consider a system consisting of the three-level device now connected through a two-level wire to the work bath, schematically shown in Figure \ref{fig:figure1}(a). The device and wire Hamiltonians read

\begin{equation}
\hat{H}_D=\omega_c\hbar|2_D\rangle\langle 2_D|+\omega_h\hbar|3_D\rangle\langle 3_D|\,,
\end{equation}

and

\begin{equation}
\hat{H}_{wire}=\omega_w\hbar|2_W\rangle\langle 2_W|\,.
\end{equation}

 The operators in the coupling terms with the baths are taken as $\hat{S}^\alpha=(\hat{S}_-^\alpha+\hat{S}_+^\alpha)$, with $\hat{S}_+^{\alpha}=\hat{S}_-^{\alpha\dagger}$ and 

\begin{equation}\label{eq:couplingbath}
\hat{S}_-^c=|1_D\rangle\langle 2_D|;\;
\hat{S}_-^h=|1_D\rangle\langle 3_D|;\;
\hat{S}_-^w=|1_W\rangle\langle 2_W|\,.
\end{equation}

The interaction between the device and the wire is described by

\begin{equation}
\hat{H}_{D,wire}=g\hbar(|3_D1_W\rangle\langle 2_D2_W|+|2_D2_W\rangle\langle 3_D1_W|)\,,
\end{equation}

where the parameter $g$ is the coupling strength. The eigenfrequencies of $\hat{H}_S=\hat{H}_{D}\,+\,\hat{H}_{D,wire}\,+\,\hat{H}_{wire}$ are $\omega_1=0$, $\omega_2=\omega_w$, $\omega_3=\omega_c$, $\omega_4=[2\omega_h-\Delta-(\Delta^2+4g^2)^{1/2}]/2$, $\omega_5=[2\omega_h-\Delta+(\Delta^2+4g^2)^{1/2}]/2$ and $\omega_6=\omega_w+\omega_h$. We have introduced the detuning $\Delta=\omega_h-\omega_c-\omega_w$. Using the procedure of section \ref{sec:theory} to determine the master equation (\ref{eq:master}), we obtain the following non-zero transition rates $W_{ij}^\alpha$ with indexes $j>i$,

\begin{equation}
\begin{array}{lll}
W_{13}^c=\Gamma_{\omega_c}^c\,, &
W_{24}^c=|c_-|^2 \,{\Gamma_{\omega_4-\omega_w}^c}\,, &
W_{25}^c=|c_+|^2 \,{\Gamma_{\omega_5-\omega_w}^c}\,, \\
W_{14}^h=|c_-'|^2\,{\Gamma_{\omega_4}^h}\,, &
W_{15}^h=|c_+'|^2\,{\Gamma_{\omega_5}^h}\,, &
W_{26}^h=\Gamma_{\omega_h}^h\,, \\
W_{12}^w=\Gamma_{\omega_w}^w\,, &
W_{34}^w=|c_-|^2\,{\Gamma_{\omega_4-\omega_c}^w}\,, &
W_{35}^w=|c_+|^2\,{\Gamma_{\omega_5-\omega_c}^w}\,, \\
W_{46}^w=|c_-'|^2\,{\Gamma_{\omega_w+\omega_h-\omega_4}^w}\,, &
W_{56}^w=|c_+'|^2\,{\Gamma_{\omega_w+\omega_h-\omega_5}^w}\,, & \\
\end{array}
\end{equation}

where the coefficients are given by

\begin{eqnarray}
c_\pm &=& \frac{[-\Delta\pm(\Delta^2+4g^2)^{1/2}]\,d_\pm}{4g(\Delta^2+4g^2)^{1/2}}\,,
\nonumber\\
c_\pm' &=& \frac{d_\pm}{2(\Delta^2+4g^2)^{1/2}}\,,
\end{eqnarray}

with $d_\pm^2=4g^2+[\Delta\pm(\Delta^2+4g^2)^{1/2}]^2$. The remaining elements can be obtained using (\ref{eq:diag}) and (\ref{eq:nondiag}). The graph representation of the master equation is shown in figure \ref{fig:figure1}(b) where we identified $38$ simple circuits using the methods described in appendix A. As each pair of vertices is connected by only one edge, the sequence of $E\leq 6$ vertices $\{i_1,i_2,\dots,i_E,i_1\}$ will denote in the following both a circuit containing these vertices and the corresponding cycle with orientation $i_1\rightarrow i_2\dots\rightarrow i_E\rightarrow i_1$. As we are interested in the system operating as a refrigerator we will focus our analysis on the steady state heat current with the cold bath. Now we will assume $\Delta=0$, for which $\omega_4=\omega_h-g$, $\omega_5=\omega_h+g$ and $|c_\pm|=|c_\pm'|^2=\frac{1}{2}$. The non-resonant case will be discussed later.

\begin{figure}[t]
\centering
\includegraphics[width=0.45\linewidth]{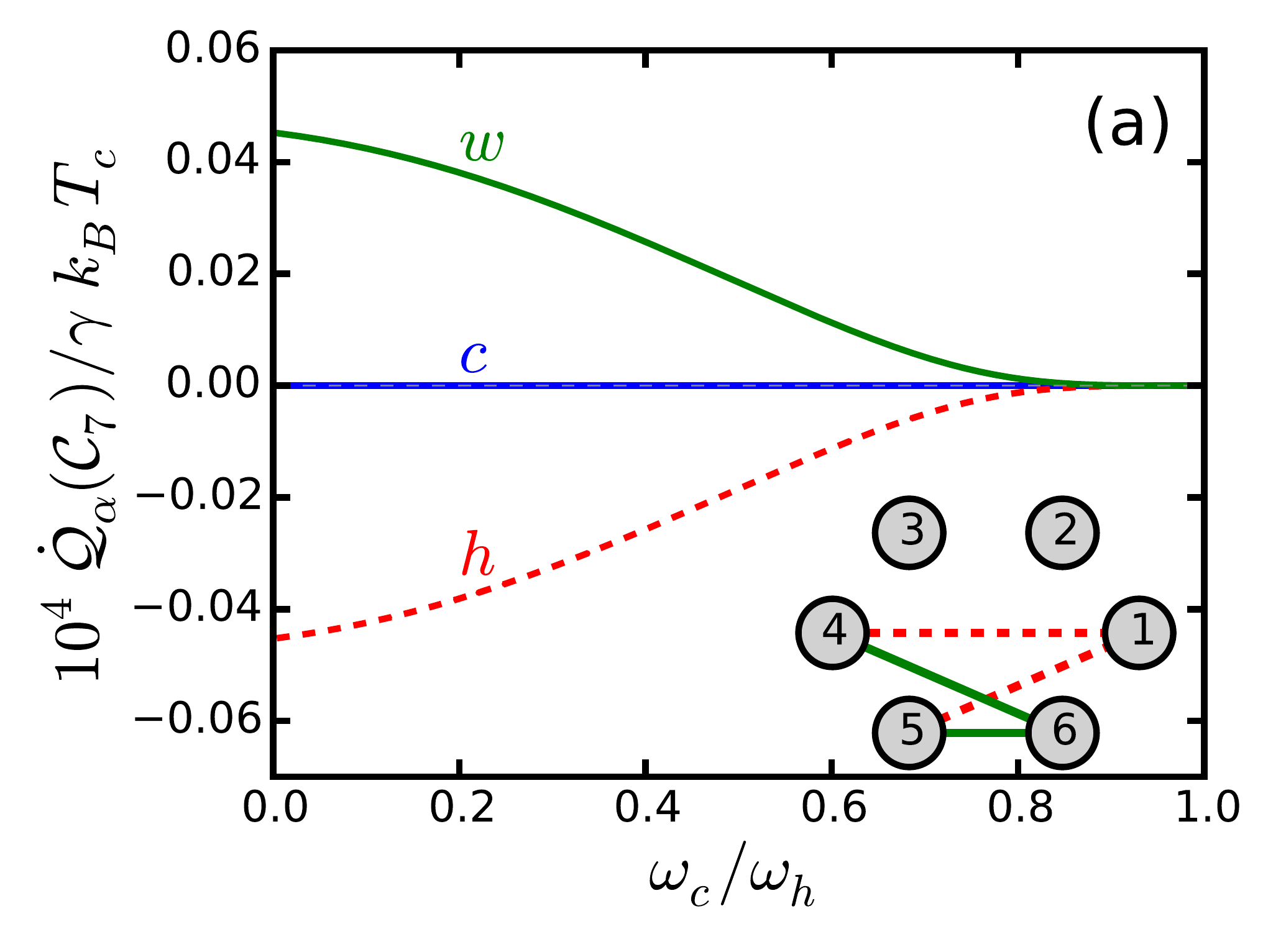}
\includegraphics[width=0.45\linewidth]{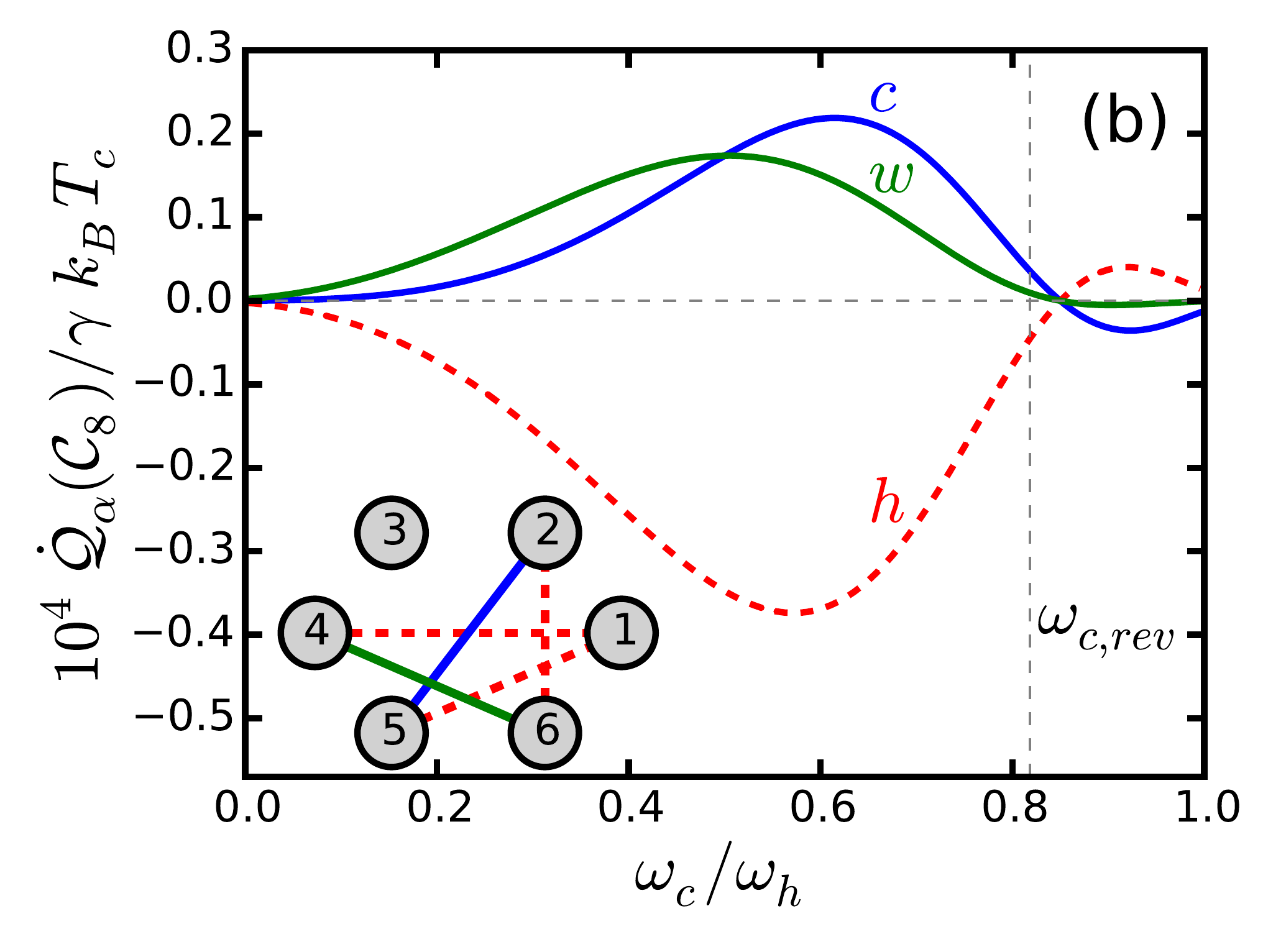}
\includegraphics[width=0.45\linewidth]{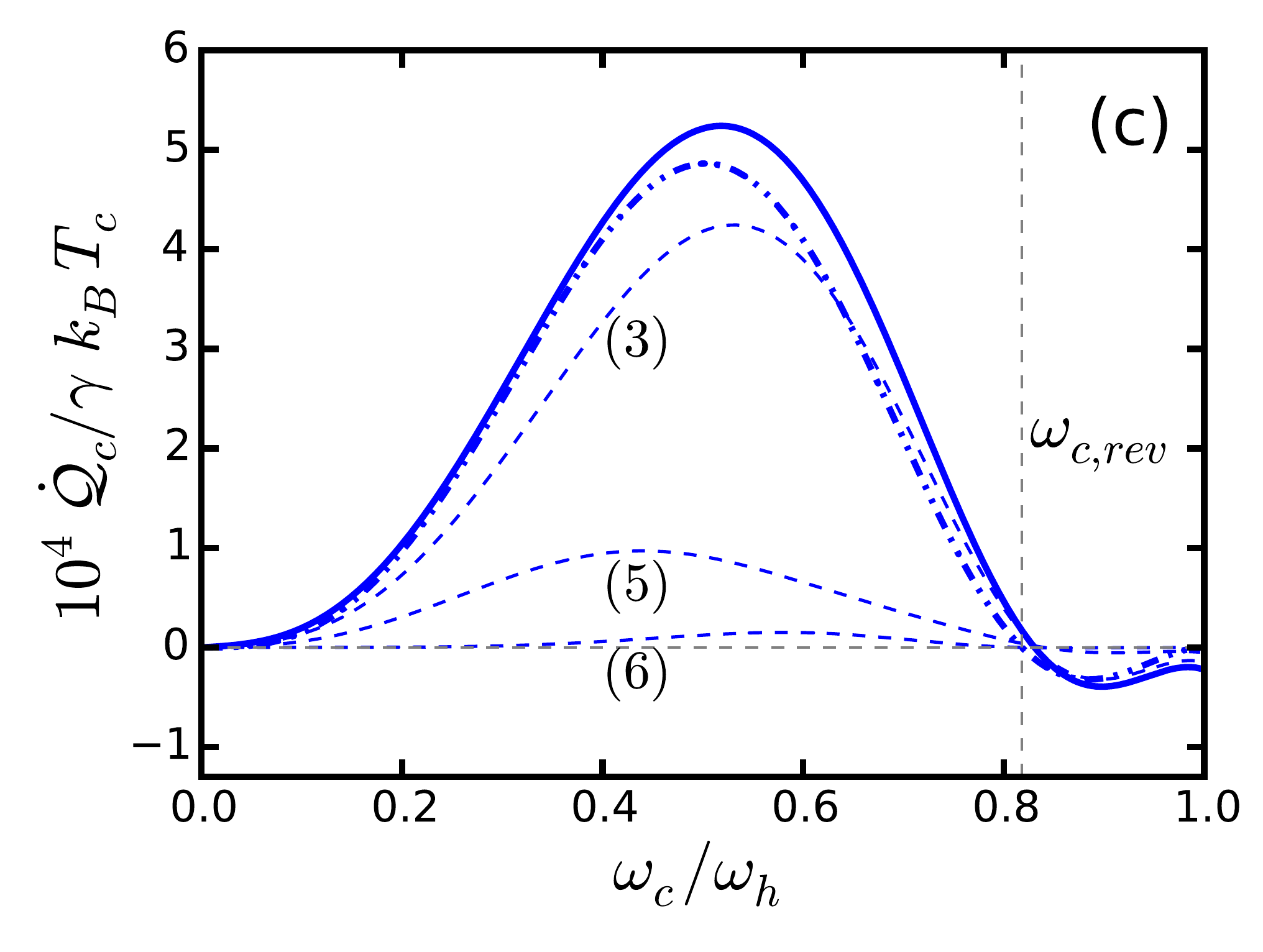}
\includegraphics[width=0.45\linewidth]{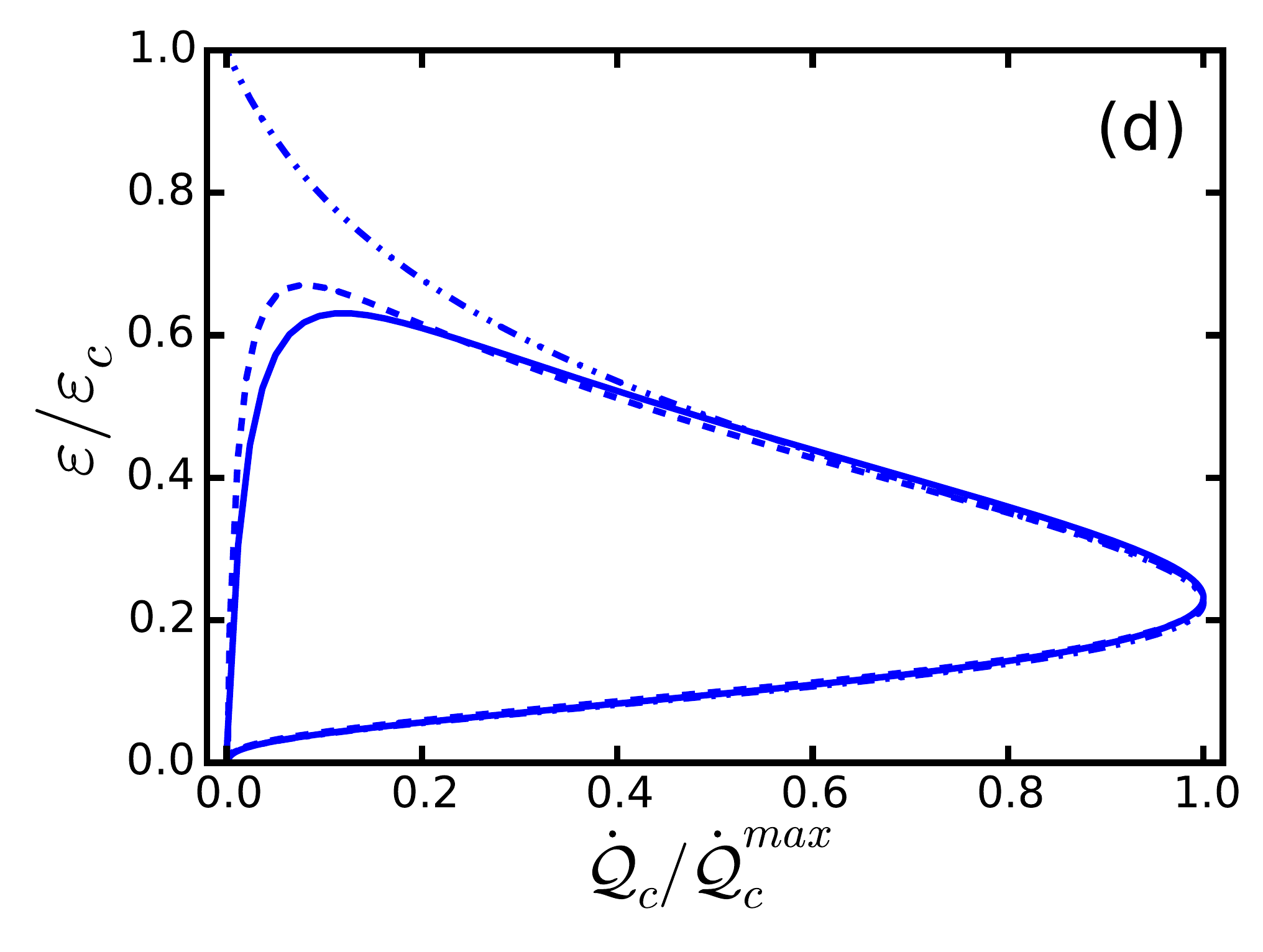}
\caption{Steady state heat currents as a function of the frequency $\omega_c$ for (a) a circuit associated with a heat leak and (b) a five-edge tricycle. The circuits are shown in the insets. (c) Heat current with the cold bath $\dot{\mathcal{Q}}_c$ for the three-level device connected through the two-level wire (solid line), the three-level directly coupled with the baths (dashed-dotted line). The labeled thin dashed lines show the contribution of tricycles with $3$, $5$ and $6$ edges. The vertical line indicates the frequency $\omega_{c,rev}$. (d) Performance characteristics obtained from (c) in the cooling range. The dashed line depicts the performance characteristic of the circuit representatives $\mathcal{C}_1$ and $\mathcal{C}_2$. Each curve has been normalized with respect to its maximum cooling rate. Parameters were chosen as $T_c=9$, $T_h=10$, $T_w=20$, $\omega_h=1$, $g=0.05$, $d_\alpha=3$ and $\gamma_\alpha=\gamma=10^{-6}$ in units for which $\hbar=k_B=\omega_0=1$.}\label{fig:figure2}
\end{figure} 

The simplest circuits in the graph are the three-edge tricycles ${\mathcal{C}}_1=\{1,3,4,1\}$, ${\mathcal{C}}_2=\{1,3,5,1\}$, ${\mathcal{C}}_3=\{1,2,5,1\}$, ${\mathcal{C}}_4=\{1,2,4,1\}$, ${\mathcal{C}}_5=\{2,5,6,2\}$ and ${\mathcal{C}}_6=\{2,4,6,2\}$, with affinities $X^c(\vec{\mathcal{C}}_{1,2})=-\omega_c\hbar/T_c$, $X^c(\vec{\mathcal{C}}_{3,5})=-(\omega_c+g)\hbar/T_c$ and $X^c(\vec{\mathcal{C}}_{4,6})=-(\omega_c-g)\hbar/T_c$. In all cases the leading term of the affinity is proportional to the frequency of the transition coupled with the cold bath when $g\ll\omega_c$. From (\ref{eq:circuitcurrent}) we find that the upper limit of the cooling window ($\dot{\mathcal{Q}}_c>0$) for the circuits $\mathcal{C}_{1}$ and $\mathcal{C}_{2}$ is given by 

\begin{equation}\label{eq:abw}
\omega_{c,rev}(\mathcal{C}_\nu)\,=\,\omega_{c,rev}\,+\,(-1)^\nu g \frac{T_c(T_w-T_h)}{T_h(T_w-T_c)}\,.
\end{equation}

A similar analysis gives $\omega_{c,rev}(\mathcal{C}_\nu)\,=\,\omega_{c,rev}\,+\,(-1)^\nu g {T_w(T_h-T_c)}/{T_h(T_w-T_c)}$ for $\nu=3,4$ and $\omega_{c,rev}(\mathcal{C}_\nu)\,=\,\omega_{c,rev}\,+\,(-1)^\nu\,g$ for $\nu=5,6$. The tricycles reach the Carnot COP when approaching to $\omega_{c,rev}(\mathcal{C}_\nu)$, but with vanishing circuit heat currents. 

We identify $10$ four-edge circuits associated with heat leaks, $4$ involving the cold and work baths, $4$ the work and hot baths and $2$ the cold and hot baths. In all cases the two non-zero affinities $X^\alpha$ are proportional to the coupling strengh $g$. An example is the circuit $\mathcal{C}_7\equiv\{1,4,6,5,1\}$ shown in figure \ref{fig:figure2}(a). In this case the affinities are $X^h(\vec{\mathcal{C}}_7)=2g\hbar/T_h$ and $X^w(\vec{\mathcal{C}}_7)=-2g\hbar/T_w$.  From the circuit heat currents (\ref{eq:circuitcurrent}), the condition $T_h<T_w$ and the relation (\ref{eq:nondiag}), one can easily determine that $\dot{\mathcal{Q}}_h(C_7)<0$ and $\dot{\mathcal{Q}}_w(\mathcal{C}_7)>0$, resulting in a direct energy transfer from the work bath to the hot bath. There is also a trivial four-edge circuit involving only the edges associated with the work bath. We found $14$ five-edge tricycles, as for example $\mathcal{C}_8=\{1,4,6,2,5,1\}$ shown in figure \ref{fig:figure2}(b). In this case $X^c(\vec{\mathcal{C}}_8)=-(\omega_c+g)\hbar/T_c$ and the circuit cooling window is given by the condition $\omega_c<\omega_{c,rev}(\mathcal{C}_8)=\omega_{c,rev}+g(2T_cT_w-T_cT_h-T_wT_h)/(T_wT_h-T_hT_c)$. Finally there are $7$ six-edge circuits, $2$ tricycles and $5$ associated to heat leaks. All the tricycles have affinities $X^c$ with a leading term proportional to $\omega_c$ and cooling windows in the interval $\omega_c-g\leq\omega_{c,rev}(\mathcal{C}_\lambda)\leq\omega_c+g$, whereas the non-zero affinities $X^\alpha$ of circuits associated with heat leaks are proportional to the coupling constant $g$.

For a given choice of the system parameters the cooling power is determined by the positive contribution of the tricycles operating in their cooling window and the negative contribution of the other tricycles and the circuits associated with heat leaks. The optimal coupling constant $g$ satisfies $\gamma_\alpha\ll g\ll\omega_c$, a regime where the contribution of the heat leaks is very small, the tricycles cooling windows approximately coincide and the rotating wave approximation is still valid. An example is shown in \ref{fig:figure2}(c). As expected, the larger contributions correspond to the tricycles with a lower number of edges. The maximum cooling rate is slightly greater and displaced to higher frequencies when the device is connected through the two-level wire. This effect is the result of the evaluation of the rate functions (\ref{eq:rates}) at the displaced frequencies $\omega_\alpha\pm g$, and increases with the bath physical dimension $d_\alpha$. However, it cannot be further exploited as a larger $g$ would increase the heat leaks, and in any case the COP would not improve.

If we examine the system performance characteristic, see figure \ref{fig:figure2}(d), a closed curve is found for the indirectly-coupled three-level device indicating the existence of additional irreversible processes: heat leaks, with small influence for small $g$, and the internal dissipation appearing when approaching the upper limit of the cooling window. The internal dissipation results from the competition of positive and negative heat currents associated with tricycles having slightly different values of $\omega_{c,rev}(\mathcal{C}_\nu)$ \cite{Correa2015} and only works for very small $\omega_c$ (where the finite heat transfer rate effects dominate) and in the interval $\omega_{c,rev}-g<\omega_c<\omega_{c,rev}-g$. These irreversible contributions can be reduced decreasing the coupling strength $g$, but cannot be avoided, making the reversible limit unattainable for the device connected through a quantum wire. 

For optimal coupling constant $g$, the main features of the system performance results from the tricycle contributions and can be described with a small number of circuit representatives. We choose as circuit representatives $\mathcal{C}_1$ and $\mathcal{C}_2$, with $\omega_{c,rev}(\mathcal{C}_{1,2})$ below and above $\omega_{c,rev}$ respectively, see (\ref{eq:abw}). Their fluxes are given by

\begin{eqnarray}
I(\vec{\mathcal{C}}_1) &=& D^{-1}\det(-\mathbf{W}|\mathcal{C}_1)\left(W_{31}^c W_{43}^w W_{14}^h-W_{13}^c W_{34}^w W_{41}^h\right)\,,
\nonumber\\
I(\vec{\mathcal{C}}_2) &=& D^{-1}\det(-\mathbf{W}|\mathcal{C}_2)\left(W_{31}^c W_{53}^w W_{15}^h-W_{13}^c W_{35}^w W_{51}^h\right)\,,
\end{eqnarray}

from which the steady state heat currents of the circuit representatives $\dot{\mathcal{Q}}_\alpha^R=\dot{\mathcal{Q}}_\alpha(\mathcal{C}_1)+\dot{\mathcal{Q}}_\alpha(\mathcal{C}_2)$ can be easily obtained. The currents $\dot{\mathcal{Q}}_\alpha^R$ incorporate the main features of the system performance such as the frequency at which the maximum cooling rate is reached and the essential irreversible processes, and  might be renormalized to account for the total heat currents \cite{Knoch2015}. Figure \ref{fig:figure2}(d) compares the performance characteristic of the system and its circuit representatives. For a more accurate description, or for larger values of $g$, a larger number of circuit representatives, including for example heat leaks, might be needed.  

The analysis of the non-resonant case leads to the same qualitative results, as the structure of the graph representation of the master equation is not modified and the same simple circuits and irreversible mechanisms are found. The main difference is the dependence of the transitions rates $W_{ij}^\alpha$ on the detuning $\Delta$. When $\gamma_\alpha\ll g\ll\Delta$, the coefficients $c_+$ and $c_-'$ vanish and three circuits associated with heat leaks dominate the heat currents: $\{1,2,4,3,1\}$ (from the work bath to the cold one) $\{1,5,6,2,1\}$ (from the work to the hot) and $\{1,3,4,2,6,5,1\}$ (from the hot to the cold).

%%%%%%%%%%%%%%%%%%%%%%%%%%%%%%%%%%%%%%%%%%

\section{Quantum three-level device coupled through a two-level system to a work source}\label{sec:worksource}

The three-level device coupled with a work source modeled by a periodic classical field is the simplest model of driven quantum thermal machines \cite{Kosloff2014}. The engine efficiency is given by $\eta=-\mathcal{P}/\dot{\mathcal{Q}}_h$ and the refrigerator COP by $\varepsilon=\dot{\mathcal{Q}}_c/\mathcal{P}$. The operating mode of the device is determined by the frequency of the transition coupled with the cold bath. When $\omega_c<\omega_{c,max}- \lambda \eta_C$, the device works as a refrigerator whereas for $\omega_c>\omega_{c,max}+\lambda \eta_C$ it works as an engine. Here we have introduced the coupling strength with the field $\lambda$ and the engine Carnot efficiency $\eta_C=1-T_c/T_h$. At the limit frequency $\omega_{c,max}=\omega_h T_c/T_h$, an idealized device ($\lambda=0$) would reach the engine Carnot efficiency or the refrigerator Carnot COP, $\varepsilon_C=T_c/(T_h-T_c)$. However, when $\omega_{c,max}-\lambda\eta_C<\omega_c<\omega_{c,max}+\lambda\eta_C$ the operating mode of the three-level device is given by the competition between the heat currents associated with the two manifold resulting from the splitting of the system energy levels due to the field interaction. The competition of those heat currents is the origin of the internal dissipation preventing the system to reach the Carnot performance in any of the two working modes \cite{Kosloff2014}. In this section we will analyze the three-level device connected to a classical driving field through the two-level wire. The system Hamiltonian is

\begin{equation}\label{eq:drivenHamiltonian}
\hat{H}\,=\,\hat{H}_{D}\,+\,\hat{H}_{D,wire}\,+\,\hat{H}_{wire}
\,+\,\hat{H}_{wire,w}(t)\,+\
\sum_{\alpha=c,h}\,\left(\hat{H}_{D,\alpha}\,+\,\hat{H}_\alpha\right)\,.
\end{equation}

All these terms were already introduced in the previous section except for $\hat{H}_{wire,w}(t)$, which describes the coupling of the two-level system with the classical field,

\begin{equation}
\hat{H}_{wire,w}(t)\,=\,\lambda\hbar\,[|2_W\rangle\langle 1_W|\exp(-i\omega_w t)\,+\,|1_W\rangle\langle 2_W|\exp(i\omega_w t)]\,.
\end{equation}

We will assume the resonant case in which the field frequency is equal to $\omega_w=\omega_h-\omega_c$. As the Hamiltonian (\ref{eq:drivenHamiltonian}) depends on time, the derivation of the quantum master equation described in section \ref{sec:theory} requires some modifications \cite{Alicki2006,Levy2012,Szczygielski2013}. Let us define the operators $\hat{H}_1=\hat{H}_D+\hat{H}_{wire}$ and

\begin{equation}
\hat{H}_2\,=\,\hat{H}_{D,wire}\,+\,
\lambda\hbar\,(|1_D2_W\rangle\langle 1_D1_W|+|2_D2_W\rangle\langle 2_D1_W|+
|3_D2_W\rangle\langle 3_D1_W|\,+\,h.c.)\,,
\end{equation}

where $h.c.$ stands for the Hermitian conjugate of the preceding terms. The eigenfrequencies of $\hat{H}_2$, $\hat{H}_2|i\rangle=\omega_i\hbar|i\rangle$, are 
$\omega_1=-\lambda$,
$\omega_2=\lambda$,
$\omega_3=-[g+(4\lambda^2+g^2)^{1/2}]/2$,
$\omega_4=[g-(4\lambda^2+g^2)^{1/2}]/2$,
$\omega_5=[-g+(4\lambda^2+g^2)^{1/2}]/2$ and
$\omega_6=[g+(4\lambda^2+g^2)^{1/2}]/2$. One can easily probe \cite{Szczygielski2013} that the propagator associated with $\hat{H}_S(t)=\hat{H}_{D}\,+\,\hat{H}_{D,wire}\,+\,\hat{H}_{wire}\,+\,\hat{H}_{wire,w}(t)$ is given by $\hat{U}_S(t)\,=\,\hat{U}_1(t)\hat{U}_2(t)$ . In the following we assume that the Lamb shifts of the energy levels of $H_S$ can be neglected. 
The coupling operators with the cold and hot baths (\ref{eq:couplingbath}) are then decomposed into

\begin{equation}
\hat{U}_S^\dagger(t)\,\hat{S}^{\alpha}\,\hat{U}_S(t)\,=\,\sum_{i=1}^{5}\sum_{j>i}\;
\hat{S}_{ij}^{\alpha}\exp[-i\,(\omega_\alpha+\omega_{ij})\,t]\,+\,
\hat{S}_{ij}^{\alpha\dagger}\exp[i\,(\omega_\alpha+\omega_{ij})\,t]
\,,
\end{equation}

where $\hat{S}_{ij}^{\alpha}=c_{ij}^\alpha|i\rangle\langle j|$ and $c_{ij}^\alpha=\langle i|\hat{S}_-^\alpha|j\rangle$. With these ingredients the LKGS generators for each bath (\ref{eq:generators}) can be obtained as the summation of the terms corresponding to the frequencies $\omega_\alpha+\omega_{ij}$, leading to the following quantum master equation in the interaction picture under the unitary transformation associated with $\hat{U}_S^\dagger$,

\begin{equation}\label{eq:qminteraction}
\frac{d}{dt} \hat{\rho}_I(t)\,=\,\sum_{\alpha=c,h}\mathcal{L}_\alpha[\hat{\rho}_I(t)]\,.
\end{equation}
\begin{figure}[h]
\centering
\includegraphics[width=0.45\linewidth]{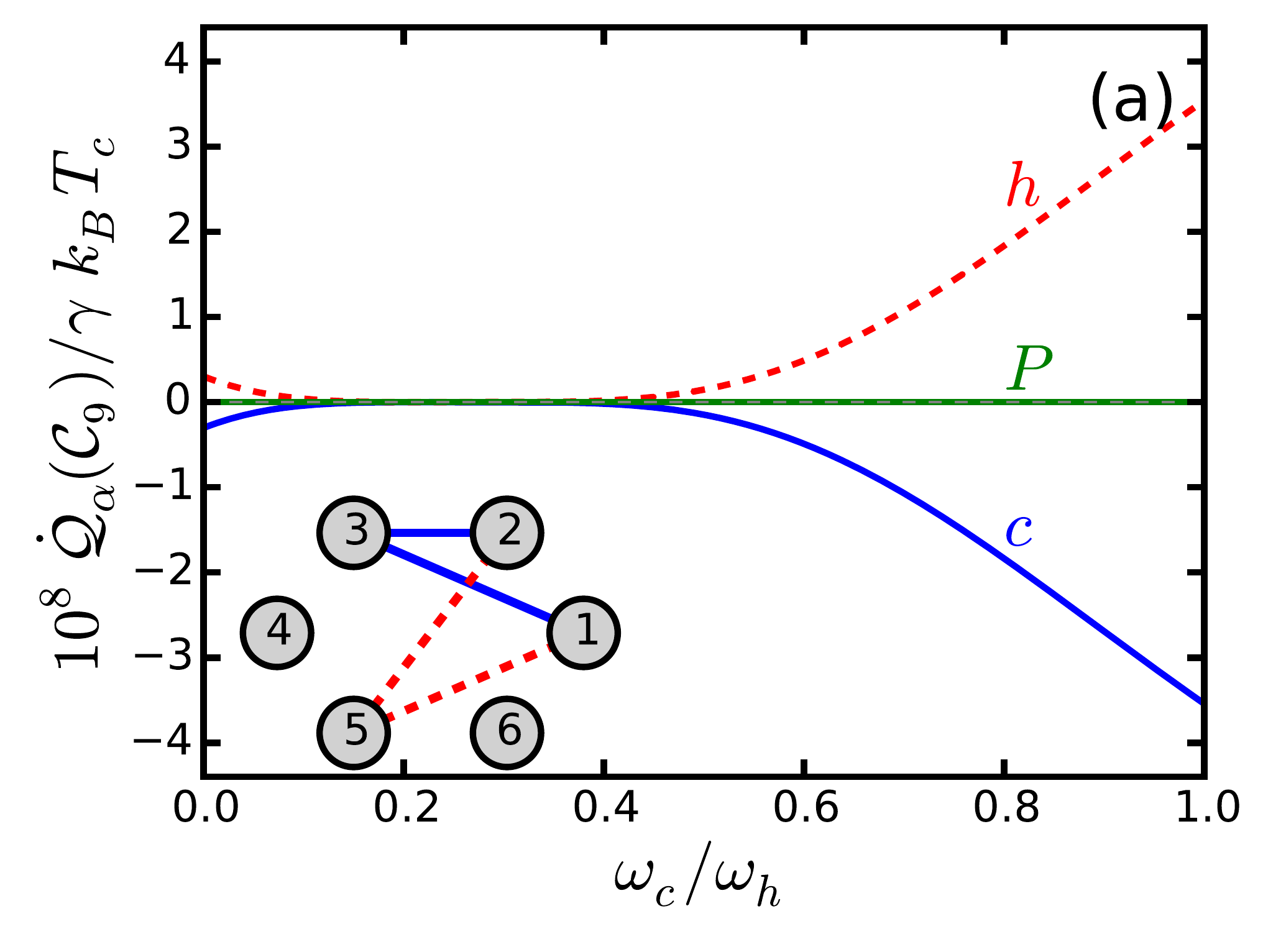}
\includegraphics[width=0.45\linewidth]{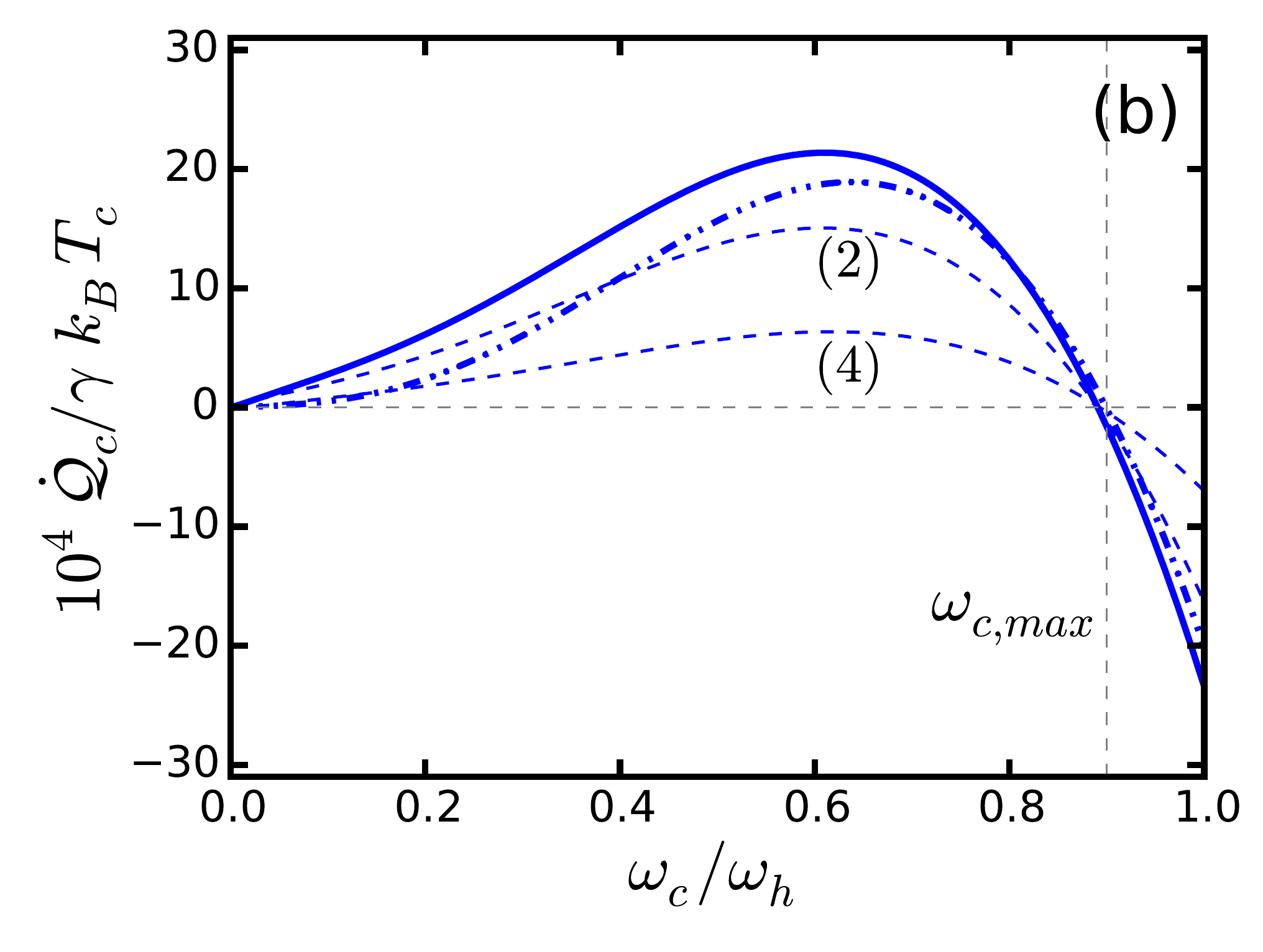}
\includegraphics[width=0.45\linewidth]{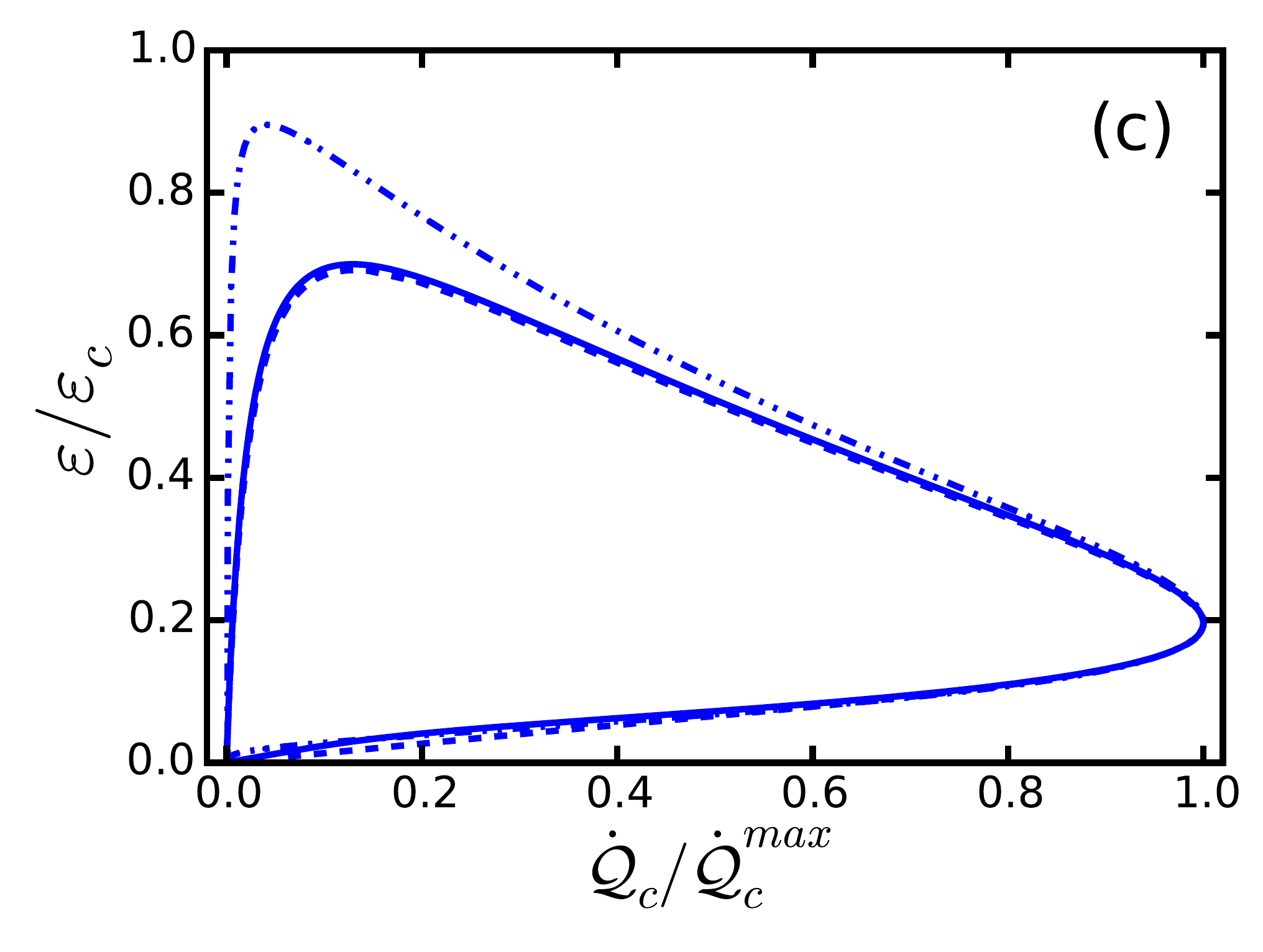}
\includegraphics[width=0.45\linewidth]{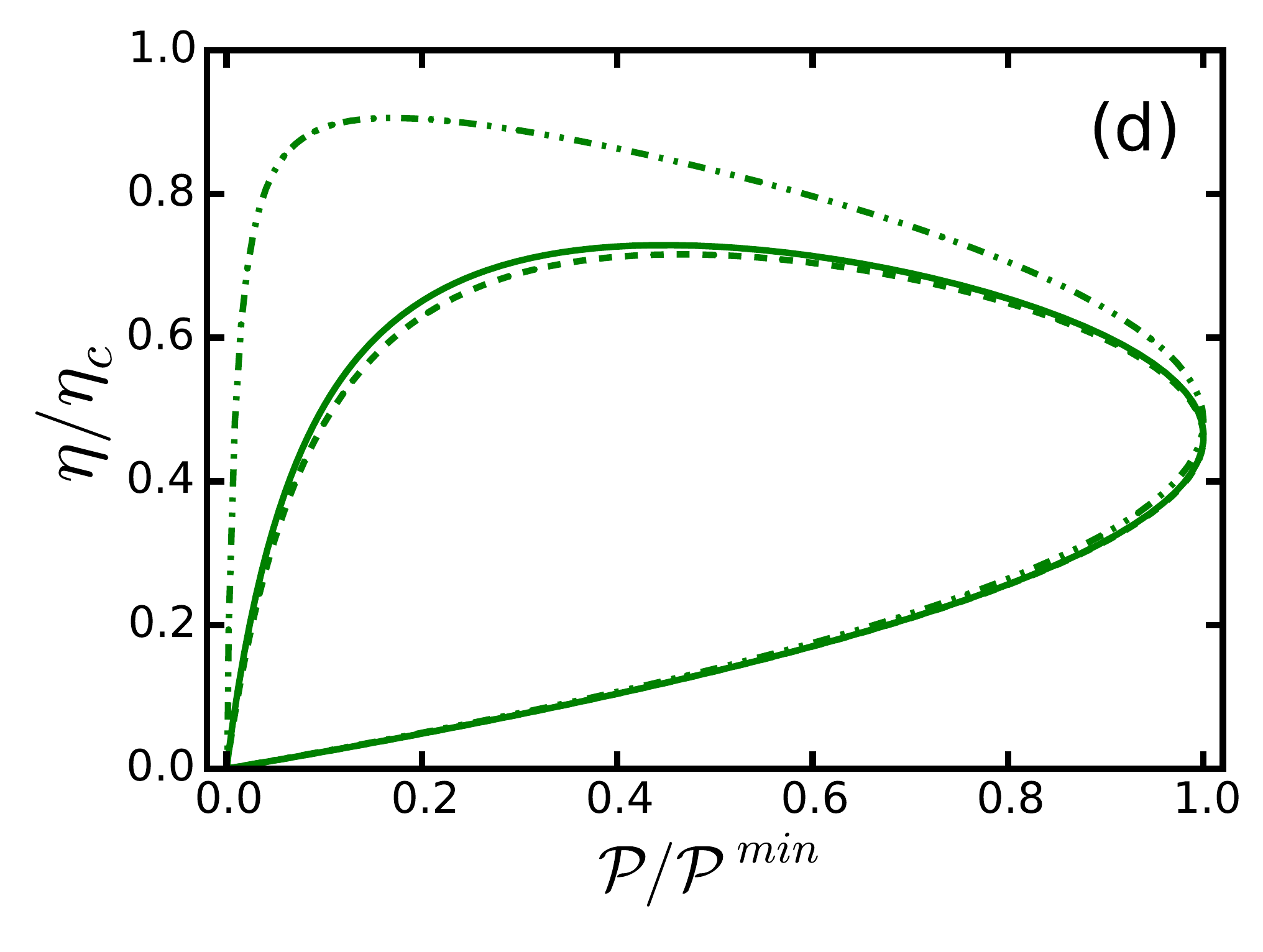}
\caption{(a) Steady state heat currents as a function of the frequency $\omega_c$ for the circuit shown in the inset. (b) Heat current with the cold bath $\dot{\mathcal{Q}}_c$ for the three-level device connected through the two-level wire (solid line) and directly coupled to the classical field (dashed-dotted line). The labeled thin dashed lines show the contribution due to tricycles with two and four edges. The vertical line indicates the frequency $\omega_{c,max}$. (c) Performance characteristics for the system when working as a refrigerator ($\omega_c<\omega_{c,max}$) and (d) as an engine ($\omega_c>\omega_{c,max}$) for the case in (b). The dashed lines depict the performance characteristics of the circuit representatives $\mathcal{C}_{1,6}$ and $\mathcal{C}_{2,3}$. We have fixed $g=0.25$, $\lambda=0.05$ and the other parameters are those in figure \ref{fig:figure2}.}\label{fig:figure3}
\end{figure} 

The steady state properties can then be derived from the diagonal part of the equation (\ref{eq:qminteraction}) in the eigenbasis of $\hat{H}_2$ , which resembles (\ref{eq:master}). Now the steady state populations $p_i$ and energy currents must be interpreted as the corresponding time-averaged quantities over a period $\tau=2\pi/\omega_w$ of the driving. The transition rates with indexes $j>i$ are given by

\begin{equation}
W_{ij}^\alpha=|c_{ij}|^2\,\Gamma_{\omega_\alpha+\omega_{ij}}^\alpha\,,
\end{equation}

where the non-zero coefficients are 

\begin{eqnarray}
|c_{13}|^2&=&|c_{26}|^2\,=\,\frac{(1-u_-)^2}{4(1+u_-^2)}\,;\;
|c_{23}|^2\,=\,|c_{16}|^2\,=\,\frac{(1+u_-)^2}{4(1+u_-^2)}\,;\;
\nonumber\\
|c_{14}|^2&=&|c_{25}|^2\,=\,\frac{(1+u_+)^2}{4(1+u_+^2)}\,;\;
|c_{24}|^2\,=\,|c_{15}|^2\,=\,\frac{(1-u_+)^2}{4(1+u_+^2)}\,,
\end{eqnarray}

with $u_\pm=2\lambda/[g\pm(4\lambda^2+g^2)^{1/2}]$. The remaining transition rates can be obtained using (\ref{eq:diag}) and (\ref{eq:nondiag}).

The graph representation of the master equation is shown in figure \ref{fig:figure1}(c). Each pair of vertices is simultaneously connected by two edges, one associated with the cold bath and the other with the hot bath. The topological structure of the graph is very simple as the states $1$ and $2$ are only coupled with $3$, $4$, $5$ and $6$. With this structure only simple circuits with two or four edges can be found. We have identified $104$ circuits, $12$ of them being trivial circuits and $92$ contributing to the steady state heat currents. The energy exchange between the system and the work source is described by the power $\mathcal{P}=\sum_\nu\mathcal{P}(\mathcal{C}_\nu)$, being $\mathcal{P}(\mathcal{C}_\nu)=-\dot{\mathcal{Q}}_c(\mathcal{C}_\nu)-\dot{\mathcal{Q}}_h(\mathcal{C}_\nu)$ the contribution of each circuit.

The simplest circuits are $8$ two-edge tricycles $\mathcal{C}_{i,j}=\{i,j,i\}$, where $i=1,2$ and $j=3,4,5,6$. In the following we will assume that the two-edge circuits are oriented choosing $i\rightarrow j$ as the edge corresponding to the cold bath. The affinities can be easily calculated to yield $X^c(\vec{\mathcal{C}}_{i,j})=-\hbar(\omega_c+\omega_j-\omega_i)/T_c$ and $X^h(\vec{\mathcal{C}}_{i,j})=\hbar(\omega_h+\omega_j-\omega_i)/T_h$. When $g,\lambda\ll\omega_{c,h}$, the leading term in the affinities is proportional to the transition frequencies. The circuit fluxes (\ref{eq:cycleflux}) are given by

\begin{equation}
I(\vec{\mathcal{C}}_{i,j})\,=\, D^{-1}\det(-\mathbf{W}|\mathcal{C}_{ij})\left(W_{ji}^c W_{ij}^h-W_{ij}^c W_{ji}^h\right)\,.
\end{equation}

With this expression the limit frequency of each circuit can be obtained imposing $I(\vec{\mathcal{C}}_{i,j})=0$ to yield

\begin{equation}\label{eq:limitfrequency}
\omega_{c,max}(\mathcal{C}_{i,j})\,=\,\omega_{c,max}-(\omega_j-\omega_i)\eta_C\,.
\end{equation}

When $\omega_{c,max}(\mathcal{C}_{i,j})$ is approached from below the circuit reaches the refrigerator Carnot COP $\varepsilon_C$, and from above the engine Carnot efficiency $\eta_C$.

We have also identified $96$ four-edge circuits $\{i,j,i',j',i\}$: $12$ trivial circuits, $60$ tricycles ($T_cX^c+T_hX^h\neq 0$) and $24$ four-edge circuits associated with heat leaks ($T_cX^c+T_hX^h=0$). The tricycles can be classified into two different groups, one involving circuits with two edges associated with each bath, for which the affinities are proportional to $2\omega_\alpha+(\omega_j+\omega_{j'})$, and the other with three edges associated to one of the baths, for which $\omega_\alpha+\omega_j-\omega_i$.  The limit frequencies for the first group are $\omega_{c,max}(\mathcal{C})=\omega_{c,max}-(\omega_j+\omega_{j'})\eta_C/2$ and for the second are given by (\ref{eq:limitfrequency}). The circuits associated with heat leaks have affinities proportional to $\omega_i-\omega_{i'}$ or $\omega_j-\omega_{j'}$. An example is the circuit ${C}_9=\{1,3,2,5,1\}$ shown in figure \ref{fig:figure3}(a) for which the affinities are $X^c(\vec{C}_3)=-2\lambda\hbar/T_c$ and $X^h(\vec{C}_3)=2\lambda\hbar/T_h$. 

The optimal coupling constants now satisfy $\gamma_\alpha\ll g,\lambda\ll\omega_c$, as the heat leaks are minimized and the energy flows are mainly determined by the contribution due to the two-edge tricycles. Figure \ref{fig:figure3}(b) shows the heat current with the cold bath for a significant value of $g$, where the contribution of the four-edge tricycles becomes relevant. As explained before for the absorption refrigerator, the device coupled through the wire reaches a larger maximum cooling rate. Although each tricycle can reach the reversible limit, their combination lead again to internal dissipation, now working in the interval $\omega_{c,max}-f(\lambda,g)\eta_C<\omega_c<\omega_{c,max}+f(\lambda,g)\eta_C$, with $f(\lambda,g)=[\lambda+g+(4\lambda^2+g^2)^{1/2}]/2$, that depends also on $g$.

We have found that the best choice of circuit representatives between the two-edge tricycles is determined by the ratio $g/\lambda$: $C_{1,4}$, $C_{2,5}$ when $g/\lambda< 1$ and $\mathcal{C}_{1,6}$, $\mathcal{C}_{2,3}$ when $g/\lambda> 1$. For $g\ll\lambda$ the system performance is well described by the uncoupled device. The performance characteristics of the system and its circuit representatives are compared in figures \ref{fig:figure3}(c) and (d) for the two operating modes. Notice that the engine maximum power output is reached at the minimum value of $\mathcal{P}$. As expected, the circuit representatives provide a more accurate description of the performance than the directly-coupled device. This approximation can be further improved by including a larger number of circuits.

%%%%%%%%%%%%%%%%%%%%%%%%%%%%%%%%%%%%%%%%%%
\section{Conclusions}\label{sec:conclusions}

In this paper we have analyzed the irreversible processes in a three-level thermal device coupled with a two-level wire connecting the system to a work bath. The coupling induces heat leaks and internal dissipation that prevents the system to reach the reversible limit. Besides, if the detuning between the transitions of {the} three-level device and the two-level wire is too large, the system stops working as an absorption refrigerator as the heat leaks dominate. We found similar results in the analysis of systems in which the wire connects either the cold or the hot bath. The additional irreversible mechanisms are proportional to the coupling constant between the device and the wire. When the wire connects the system with a work source, the coupling induces heat leaks and modifies the frequency interval where internal dissipation appears. The optimal values of the coupling constants are such that $\gamma_\alpha\ll g,\lambda\ll\omega_\alpha$, which minimize the heat leaks and the interval where the internal dissipation works. The system performance can be well described by just considering two circuit representatives. This description may be improved incorporating additional circuits and renormalizing their contribution to the heat currents \cite{Knoch2015}.

Our results can be generalized to wires composed of a chain of two-level systems. The graph representation of the master equation will have a larger number of vertices and edges, exponentially increasing the number of simple circuits. However, the graph topological structure will be similar. For example let us consider a chain of $n$ two-level systems connecting the device with a classical periodic field. The corresponding graph resembles the one in figure \ref{fig:figure1}(c), with $2^n$ states connected to $2^{n+1}$ states by two edges associated with the cold and the hot baths. Again the most important contribution to the heat currents comes from the $2^{2n+1}$ two-edge circuits. Now the transition rates will depend on the frequencies $\omega_\alpha+f_{ij}(\lambda,g,n)$ and the affinities $X^\alpha$ of circuits associated with heat leaks will be proportional to $f_{ij}$. In general the function $f_{ij}$, and therefore the frequency interval subjected to internal dissipation, increases with $n$. However, for coupling constants $g$ and $\lambda$ small enough, the heat leaks can be neglected. This same analysis applies when the system is coupled with a work bath. The only limitation in both cases is that $g$ must be much larger than the coupling constants with the bath $\gamma_\alpha$ for the master equation to be valid. 

The irreversible processes analyzed are the result of the new decay channels due to the additional coupling term. Therefore they are expected in any quantum device connected through quantum wires to the environments. To avoid them would require reservoir engineering techniques \cite{Correa2015} hindered by the complicated system spectrum as the number of states is increased.
{In this work we have focused on the stationary regime. Although coherences and populations associated with the system Hamiltonian become decoupled during the evolution, the wire may introduce new effects in the transient regime} \cite{Brask2015}, {related to the time-dependent terms in the thermodynamic fluxes. These effects could be explored in future work.}

%%%%%%%%%%%%%%%%%%%%%%%%%%%%%%%%%%%%%%%%%%

\acknowledgments{\textbf{Acknowledgments:} We thank A. Ruiz and L. A. Correa for careful reading and commenting on the manuscript. J.O.G acknowledges a FPU fellowship from Spanish MECD. Financial support by the Spanish MINECO (FIS2013-41352-P) and COST Action MP1209 is gratefully acknowledged.}

%%%%%%%%%%%%%%%%%%%%%%%%%%%%%%%%%%%%%%%%%%

\textbf{Author Contributions:} J.O.G. performed the model calculations. J.O.G., D.A. and J.P.P. conceived the study and contributed equally in the mathematical derivations, the discussion of the results and the manuscript writing.

%%%%%%%%%%%%%%%%%%%%%%%%%%%%%%%%%%%%%%%%%%

\textbf{Conflicts of Interest:} The authors declare no conflict of interest.

%%%%%%%%%%%%%%%%%%%%%%%%%%%%%%%%%%%%%%%%%%
%% optional
\appendix %%MDPI internal note: new layout%%

\section*{\noindent Appendix A}\vspace{6pt} %%MDPI internal note: new layout%%
\label{sec:appendix}

\begin{figure}[h]
\centering
\includegraphics[width=0.9\linewidth]{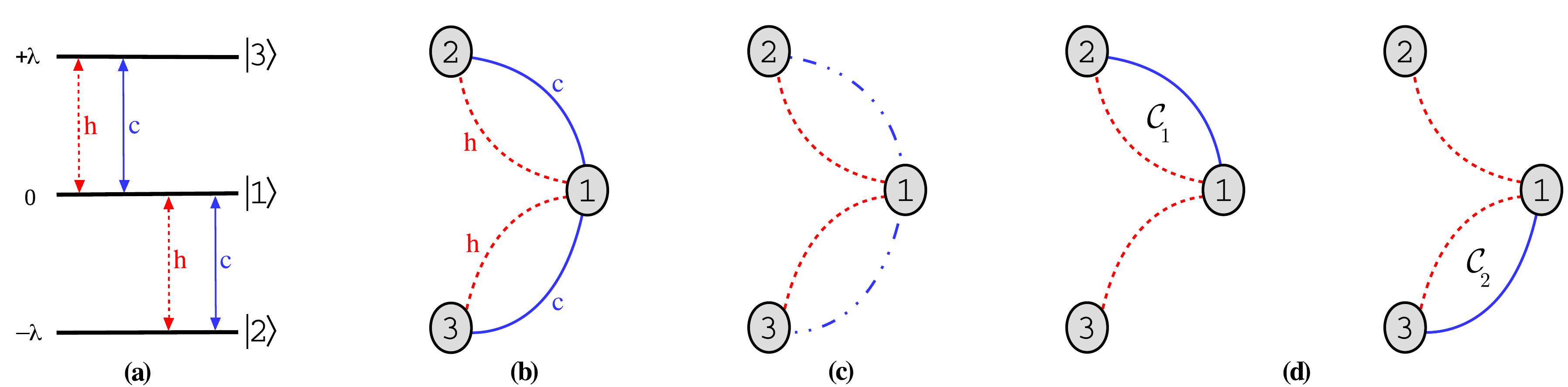}
\caption{(a) Schematic representation of the transitions due to the coupling with the cold and hot baths, labeled by $c$ and $h$ respectively, in a driven three-level system. (b) In the graph $\mathcal{G}$ the vertices represent the three states and the undirected edges the transitions connecting them. One of the graph maximal trees $\mathcal{T}$ is shown in (c), for which the chords (dashed-dotted lines) are the edges corresponding to the cold transitions. (d) A set of fundamental circuits $\{\mathcal{C}_1$, $\mathcal{C}_2\}$ is then identified by adding each chord to the maximal tree. These circuits are equivalent to the two manifolds operating in the three-level amplifier \cite{Levy2014}.}\label{fig:figureA1}
\end{figure} 

In this appendix we describe how to obtain all the circuits corresponding to a graph. This simple procedure is illustrated with the three-level system in contact with a cold and a hot unstructured bosonic baths and directly coupled with a periodic classical field. The total Hamiltonian is

\begin{equation}
\hat{H}\,=\,\hat{H}_{D}(t)\,+\,\sum_{\alpha=c,h}\,\hat{H}_{D,\alpha}\,+\,\hat{H}_\alpha\,,
\end{equation}

where 

\begin{equation}
\hat{H}_{D}(t)\,=\,\omega_c \hbar |2_D\rangle\langle 2_D|+\omega_h \hbar |3_D\rangle\langle 3_D|
+\lambda  \hbar(|3_D\rangle\langle 2_D| \exp[-i(\omega_h-\omega_c)t]+h.c.)\,.
\end{equation}

The bath Hamiltonians $\hat{H}_\alpha$ and the coupling terms $\hat{H}_{D,\alpha}$ are described in section \ref{sec:bathsource}. The quantum master equation for this system can be obtained using the procedure described in section \ref{sec:worksource} and the result can be found for example in \cite{Kosloff2014,Correa2014b}. For convenience we will use the eigenbasis $|1\rangle\equiv|1_D\rangle$, $|2\rangle\equiv2^{-1/2}(3_D\rangle-|2_D\rangle)$ and $|3\rangle\equiv2^{-1/2}(|3_D\rangle+|2_D\rangle)$ of $\hat{H}_2=\lambda\hbar(|3_D\rangle\langle 2_D|+|2_D\rangle\langle 3_D|)$. Figure \ref{fig:figureA1}(a) shows an schematic representation of the system transitions assisted by the baths. The non-diagonal elements of the rate matrix $\mathbf{W}$ are $W_{ij}=W_{ij}^c+W_{ij}^h$ with

\begin{equation}
W_{12}^\alpha=\frac{\Gamma^\alpha_{\omega_\alpha-\lambda}}{2};\; W_{13}^\alpha=\frac{\Gamma^c_{\omega_\alpha+\lambda}}{2};\;
W_{21}^\alpha=\frac{\Gamma^\alpha_{-(\omega_\alpha-\lambda)}}{2};\; W_{31}^\alpha=\frac{\Gamma^\alpha_{-(\omega_\alpha+\lambda)}}{2};\;
W_{23}^\alpha=W_{32}^\alpha=0\,.
\end{equation}

The functions $\Gamma^\alpha_{\pm\omega}$ can be calculated using (\ref{eq:rates}). The graph $\mathcal{G}$ associated with this system have $|V|=3$ vertices and $|E|=4$ edges, see Figure \ref{fig:figureA1}(b).

The procedure to determine the simple circuits is based on the identification of a maximal tree of $\mathcal{G}$ and its chords. A maximal tree $\mathcal{T}$ is a subgraph with $|V|-1$ edges connecting all the vertices but without forming any closed path. In principle many different maximal trees can be found on a given graph, but for this procedure it is sufficient to select any of them as the final result is independent of this choice. A chord of a maximal tree is one of $|E|-|V|+1$ edges which are not part of it. An example of maximal tree and its chords is shown in Figure \ref{fig:figureA1}(c).

A {\it fundamental set} of simple circuits \cite{Schnakenberg1976} can be found adding each chord to the maximal tree, as shown in Figure \ref{fig:figureA1}(d). The number of fundamental circuits equals the number of chords, $|E|-|V|+1$. Only for some systems, as our example, the fundamental set contains all the possible simple circuits. Otherwise, the remaining circuits can be obtained by the linear combination of the elements of the fundamental set 

\begin{equation}\label{eq:circuitcombination}
r_1 C_1 \oplus r_2 C_2 \oplus \dots \oplus r_{|E|-|V|+1} C_{|E|-|V|+1}\,, 
\end{equation}

with $r_\lambda=0$ or $1$. The relation $C_\lambda \oplus C_{\lambda'}$ gives a new subgraph that contains all the edges of $C_\lambda$ and $C_{\lambda'}$ which do not simultaneously belong to $C_\lambda$ and $C_{\lambda'}$ \cite{Schnakenberg1976}. The result of each possible linear combination (\ref{eq:circuitcombination}) is considered only when it generates a new simple circuit.

In summary, a simple procedure to obtain the full set of circuits reads as
\begin{itemize}
\item[(i)] Select a maximal tree $\mathcal{T}$ of $\mathcal{G}$ and identify its chords. 

\item[(ii)] Find a fundamental set of circuits adding each chord to the maximal tree.

\item[(iii)] Obtain the remaining circuits by the linear combination of the circuits in the fundamental set.
\end{itemize}

An alternative procedure consists in identifying all the maximal trees of $\mathcal{G}$ and generate the fundamental set of circuits associated with each one. The set of all simple circuits is then the union of all the fundamental sets. For complex graphs the number of circuits might be very large and more efficient standard algorithms \cite{Johnson1975} can be used.

%=================================================================
% References: Variant A
%=================================================================
% Back Matter (References and Notes)
%----------------------------------------------------------
% Style and layout of the references
\bibliographystyle{mdpi}
\renewcommand\bibname{References}
%%MDPI internal note: new layout%% redefinition removed

\end{document}